\documentclass[12pt]{article}
\usepackage{graphicx,subfig,rotate,color}
\usepackage{amsmath,amssymb,latexsym}
\newcommand{\email}{E-mail address: }
\newcommand{\tev}{{TeV~}}

\begin{document}
\begin{center}
{\large{\bf Probing Randall-Sundram Model using triphotons at the LHC}}
\vskip 25pt
{\bf David Atwood\footnote{\tt\email atwood@iastate.edu} and Sudhir Kumar Gupta\footnote{\tt\email skgupta@iastate.edu}} \\\vskip 10pt
{\em Department of Physics \& Astronomy\\ Iowa State University, Ames, IA 50010 USA}
\date{16 December 2008}
\\[.2cm]
\normalsize
\end{center}
\vskip 25pt

\begin{abstract} 
We investigate triphoton signals of the Randall-Sundram model at the 
Large Hadron Collider. Such a signal can be an important probe to the RS 
model as these are relatively cleaner from the hadronic activity and 
bear significant rate. The corresponding standard model background has 
also been studied in detail. We also show that a clear graviton 
reconstruction is possible in such signal.
\end{abstract}
\newpage

\section{Introduction}

Phenomenology of models based on extra spatial dimensions~\cite{reved} 
is quite popular now. Besides offering a solution to the hierarchy 
problem of the Standard Model~\cite{c1nat} these models allow for the 
low-energy unification of the gauge couplings~\cite{Pomarol:2000hp}, 
provide a rich \tev scale new physics phenomenology, existence of 
gravity at the \tev scale and may even offer candidate(s) for the cold 
dark matter of the universe~\cite{Agashe:2004bm}.

In the simplest string theory inspired extension of the standard model 
(SM) based on one extra spatial dimension, originally proposed by 
Randell-Sundram (RS)~\cite{Randall:1999vf,Randall:1999ee}, gravitons are 
the only propagating particles in the bulk. Such gravitons will 
therefore have Kaluza-Klein (KK) excitations which will appear in 
experiments as a widely separated resonances. This contrasts with the KK 
spectrum of models with compact extra dimensions such as 
ADD~\cite{add_paper} where there are a very large number of closely 
spaced graviton modes.

The coupling of these graviton excitations to the SM is through the 4-d 
reduced Planck mass $\bar{M}_P$, which may be on the TeV scale, rather 
than the inaccessibly large Planck mass at $10^{19}$ GeV. The couplings 
to Standard Model particles are therefore be proportional to $\sim 
1/\bar{M}_P$, thereby allowing graviton excitations to decay into all 
the SM particles including a fermion pair or a pair of gauge bosons. At 
TeV scale energies, when such graviton excitations are produced, this 
variety of possible decay modes will give rise to vast phenomenology at 
the \tev scale; $\Lambda = \bar{M}_P~e^{-\kappa r_c \pi}$, with 
$e^{-\kappa r_c \pi}$ as a warped factor which arise due to the 
compactification of the extra dimension on a circle with radius $r_c$. 
The factor $\pi$ is due to the fact that SM in located on the circle at 
$\phi=\pi$ and $\kappa$ is the curvature parameter.

%
%
%
%

Although lots of variations of the RS model have been proposed over the 
years~\cite{Csaki:1999mz} and their phenomenology~\cite{rspheno} has 
been studied in detail, in this paper we will consider the original 
scenario. In particular, we assume that the whole the SM is localized on 
the \tev brane, so that the mass of gravitons is given by $m_{G_n}= 
x_n~\kappa~e^{-\kappa r_c \pi}=x_n(\kappa/\bar{M}_P)~\Lambda$, where 
$x_n$ are the roots of the first-order Bessel function. In order to be 
useful in the resolution of the hierarchy problem and keep gravity weak 
enough to be treated perturbatively, $\kappa/\bar{M}_P$ should lie in 
the range $0.001 <\kappa/\bar{M}_P < 0.1$.

The focus of this paper will be the distinctive triphoton signature at 
the Large Hadron Collider (LHC) produced by the RS model and other 
models like it. The importance of such a signature lies in the fact that 
this signature is experimentally clean and a distinctive signature for 
models of this type. We will also discuss graviton mass reconstruction 
and the angular distribution of the graviton decay which 
which are important tools for characterizing the 
physics which produces the triphoton signal.

The paper is organized as follow: In section 2, we discuss the graviton 
production in association with a photon and its decays into the SM 
particles. Section 3 focuses on the numerical analysis of signal and 
background as well as graviton reconstruction in detail. Finally, in 
section 4, we summarize our findings.

\section{Graviton Production and Decay}

Triphoton signal in the RS model will arise due to the associated 
production of a on shell graviton with a photon while the graviton 
subsiquently decays into an additional photon pair. In this section we 
discuss the production process as well as the various other dominant two 
body decays of the graviton.

The parton-level matrix-element for the production process $q{\bar 
q} \to \gamma G$ as calculated in~\cite{addsk} is,

\begin{eqnarray}
 {\left|{\cal M}\right|^2} &=&N_c Q^2_f 
\left(\frac{3 \pi \alpha}{2}\right) 
\left(\frac{\hat s}{{\bar{M}_P}^2}\right) F(\eta, \zeta),\\
F(\zeta, \eta) &=& f(\zeta) + f(\eta) 
+ \zeta \eta \left(
\frac{1}{\zeta}\phi(\zeta) + 
\frac{1}{\eta }\phi(\eta)  
\right )
\end{eqnarray}
with,
\\ 
$$f(\zeta) = 1-\zeta^2+\zeta^3,$$
$$\phi(\zeta) = \zeta(2-11\zeta+4\zeta^2),$$
\\ 
where, 
$\zeta = \frac{\hat t}{\hat s}$
and 
$\eta = \frac{\hat u}{\hat s}$. 
$N_c$ is the  number of colors, 
$N_c = 3$ and $Q_f$ is the quark charge, $Q_f = + 2/3, 1/3 $ for 
up and down type quarks respectively. 

The kinematics of this process implies that $m_{G}^2 = {\hat s}^2 + 
{\hat t}^2 + {\hat u}^2$. Note that this cross-section is symmetric with 
respect to the interchange of $\hat t$ and $\hat u$.

\begin{figure}
\centerline{
\includegraphics[angle=-90, width=0.6\textwidth]{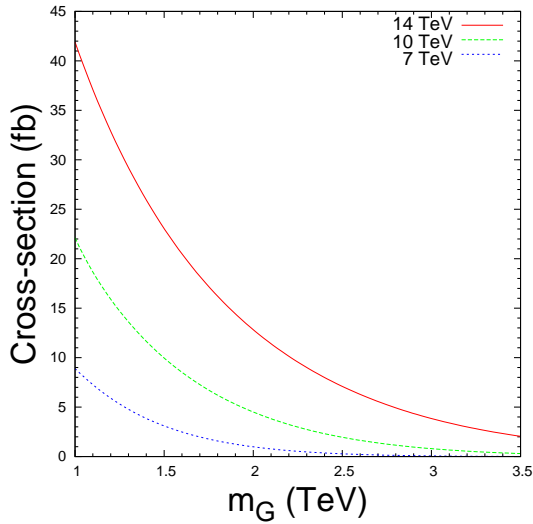}}
\caption[Production] {\small\sf Cross-section for associated production 
of a Graviton with a photon at the LHC for center-of-mass energy $\sqrt 
{s} =7$ TeV, $10$ TeV and $14$ TeV.}
\label{fig:gprd}
\end{figure}

The LHC production cross-section for this process is presented in 
Figure~\ref{fig:gprd}. We use a wide range of graviton mass well above 
the Tevatron bounds~\cite{Abazov:2007ra} for three different sets of 
center-of-mass energy, $\sqrt {s}$ as 7 TeV, 10 TeV, and, 14 TeV 
respectively.  We use CTEQ6L-1\cite{Lai:1994bb} parton densities at $Q = 
\sqrt{\hat s}$ , and the renormalization and factorization scales are 
set as, $\mu_R = Q = \mu_F$.

From the Figure~\ref{fig:gprd}, it is quite clear that due to the low 
cross-section, it is hard to observe such productions with the early LHC 
data with an integrated luminosity of 100 pb$^{-1}$ and $\sqrt {s} =10$ 
TeV. For instance only 2 events would be produced if $m_G=1$ TeV; at the 
higher energy $\sqrt {s} = 14$ TeV, the number of events is roughly 
doubled.

\begin{figure}[ht]
\begin{center}
\includegraphics[width=3.5in,height=3in]{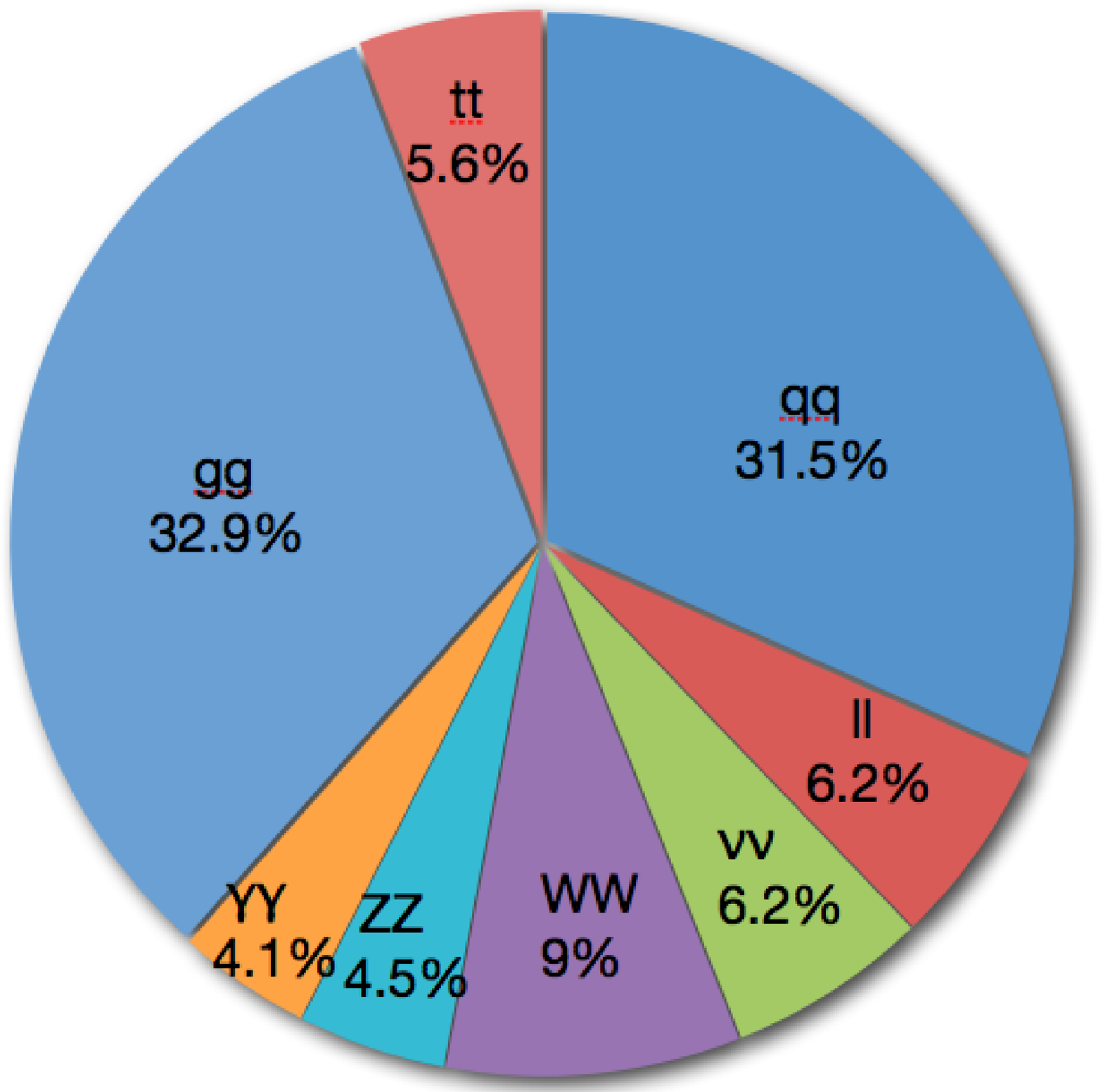}
\caption[Graviton Decay]{\small\sf Branching fractions for Graviton 
decays in to standard model particles. $m_G = 1$ \tev has been 
assumed in this chart.}
\label{fig:gdec}
\end{center}
\end{figure}

In Figure~\ref{fig:gdec}, we present the branching fractions of graviton 
to various two-body SM mode. The dominant mode is the dijet channel with 
a branching ratio $B(\longrightarrow gg, q_i\bar{ q}_i) \sim 64 \%$ 
(where, $i=u,d,s,c,b$). The fraction for the diphoton mode is smaller, 
$\sim 4\%$, but it is important as this will lead to a clean signature 
at the LHC.


\section{Triphotons at the LHC}

The production of a graviton associated with a photon occurs in the high 
x-region, so generally the associated photon as well as the photons 
produced in the graviton decay will carry high transverse 
momentum. Large transverse momentum cuts on the triphoton signal will 
therefore be helpful in selecting graviton events and rejecting SM 
backgrounds where photons tend to have low transverse momentum.

\begin{figure}[htb]
\centerline{
\includegraphics[angle=-90, width=0.6\textwidth]{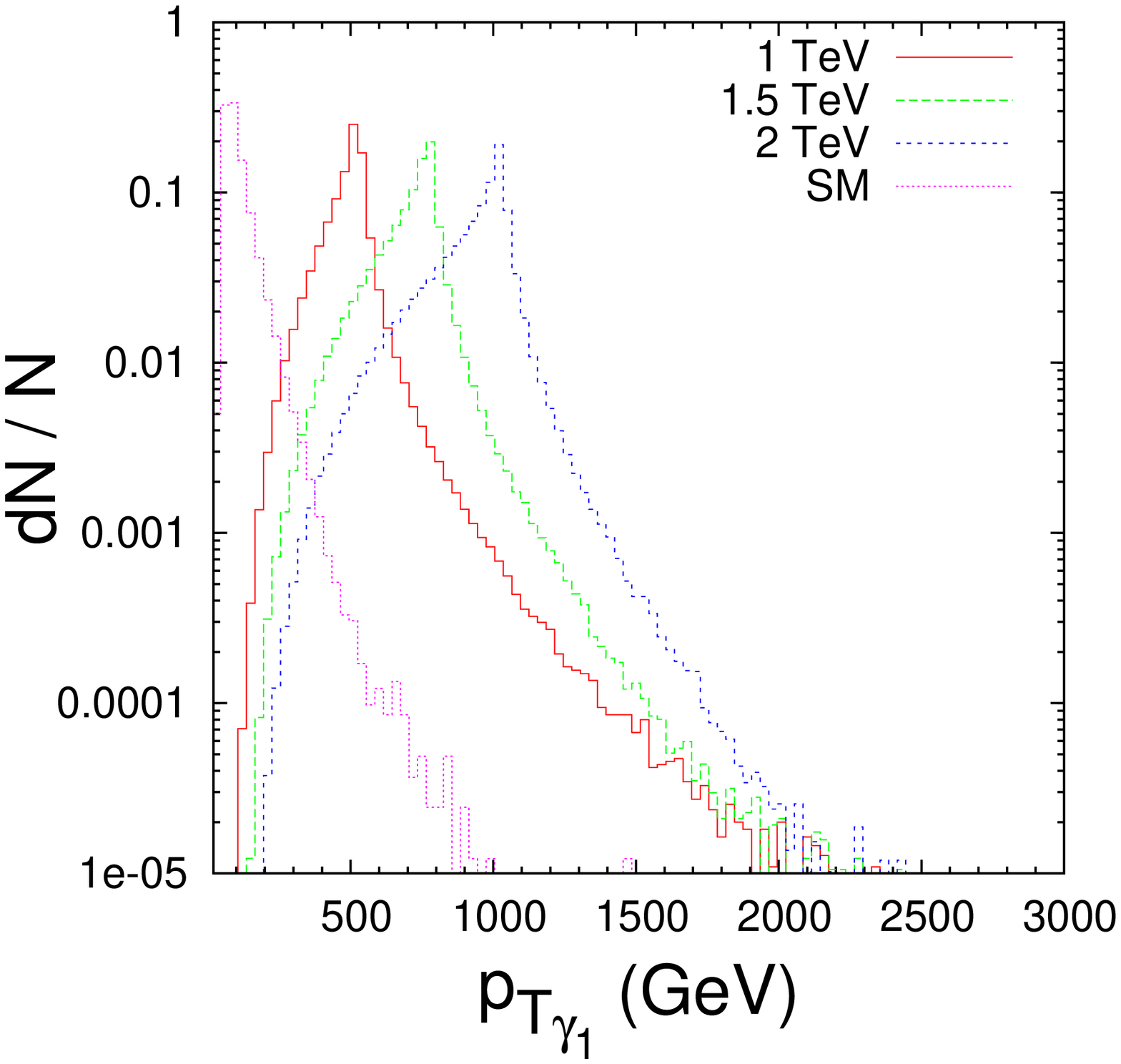}
\includegraphics[angle=-90, width=0.6\textwidth]{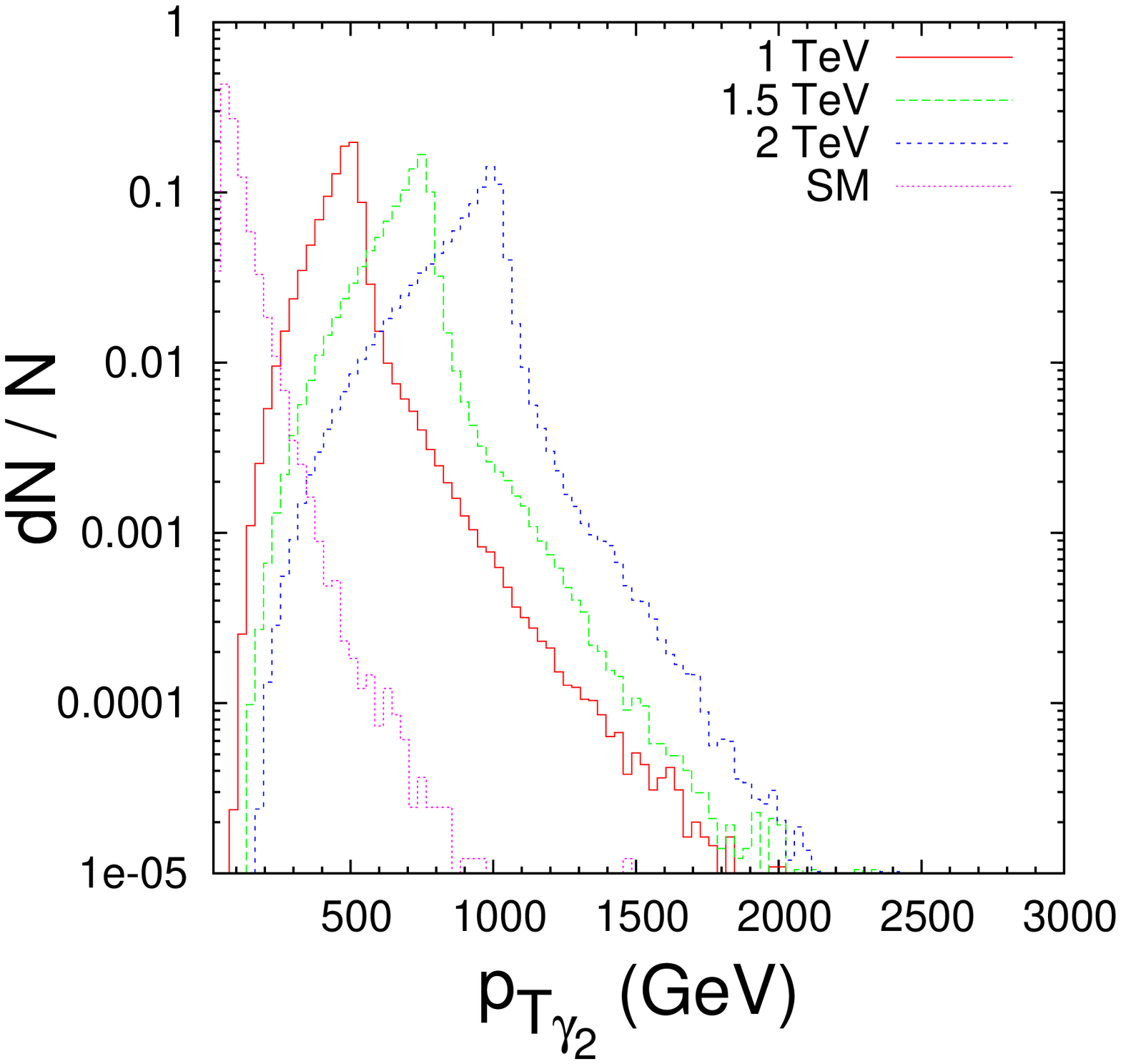}}
\centerline{\includegraphics[angle=-90, width=0.6\textwidth]{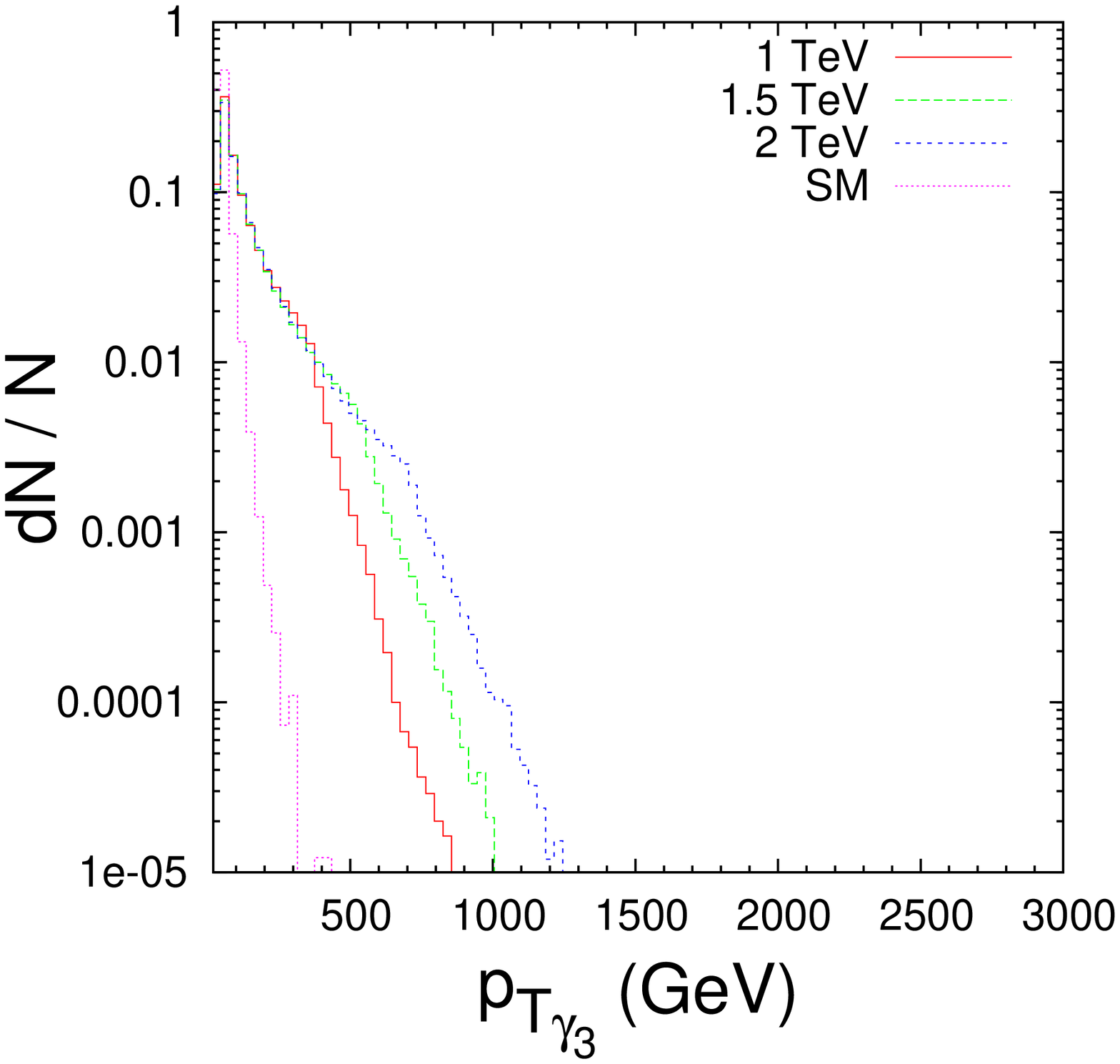}}
\caption{\small\sf $p_T$ ordered distributions of the photons. We assume 
$P_{T_{\gamma_1}} > P_{T_{\gamma_2}} > P_{T_{\gamma_3}}$ in this figure.}
\label{fig:ptg}
\end{figure}

we use {\tt MADGRAPH}~\cite{Maltoni:2002qb, Hagiwara:2008jb} to produce 
signal events with a photon and a graviton. Later, these events are 
interfaced to {\tt PYTHIA}~\cite{Sjostrand:2006za} for the analysis 
purpose. Decay of gravtiton is done using the decay table in {\tt 
PYTHIA}. Braching fractions for different decay modes for $m_G =1$ TeV are 
shown in Fig.~\ref{fig:gdec}.

Before selecting our event samples, we order the photons on the basis 
of their transverse momentum i.e., $p_{T_{\gamma_1}} > p_{T_{\gamma_2}} 
>p_{T_{\gamma_3}}$.

In order to analyze the actual event rate expected to observe at the 
LHC, we first employ the following three basic cuts:

\begin{itemize}
\item $p_{T_{\gamma_{1,2,3}}} > 25$ GeV and $|\eta_{\gamma_{1,2,3}}|< 2.7$,

\item Photon-photon separation, 
$\Delta R_{\gamma_i\gamma_j} = \sqrt{{(\eta_i - \eta_j)}^2 
+ {(\phi_i - \phi_j)}^2} > 0.2$; $i,j=1,2,3$, $i\neq j$, $\eta_i$ 
being psuedo-rapidity of photon 
$i$ and is defined as $\eta_i = -\ln(\tan\theta_i/2)$.

\item Missing transverse energy, $E_T{\!\!\!\!\!\!/\!\!\ }~ < 30 ~{\rm GeV}$.
\end{itemize}

\subsection{SM Background}
\begin{figure}
\centerline{
\label{fig:spt1}\includegraphics[angle=-90, width=0.6\textwidth]{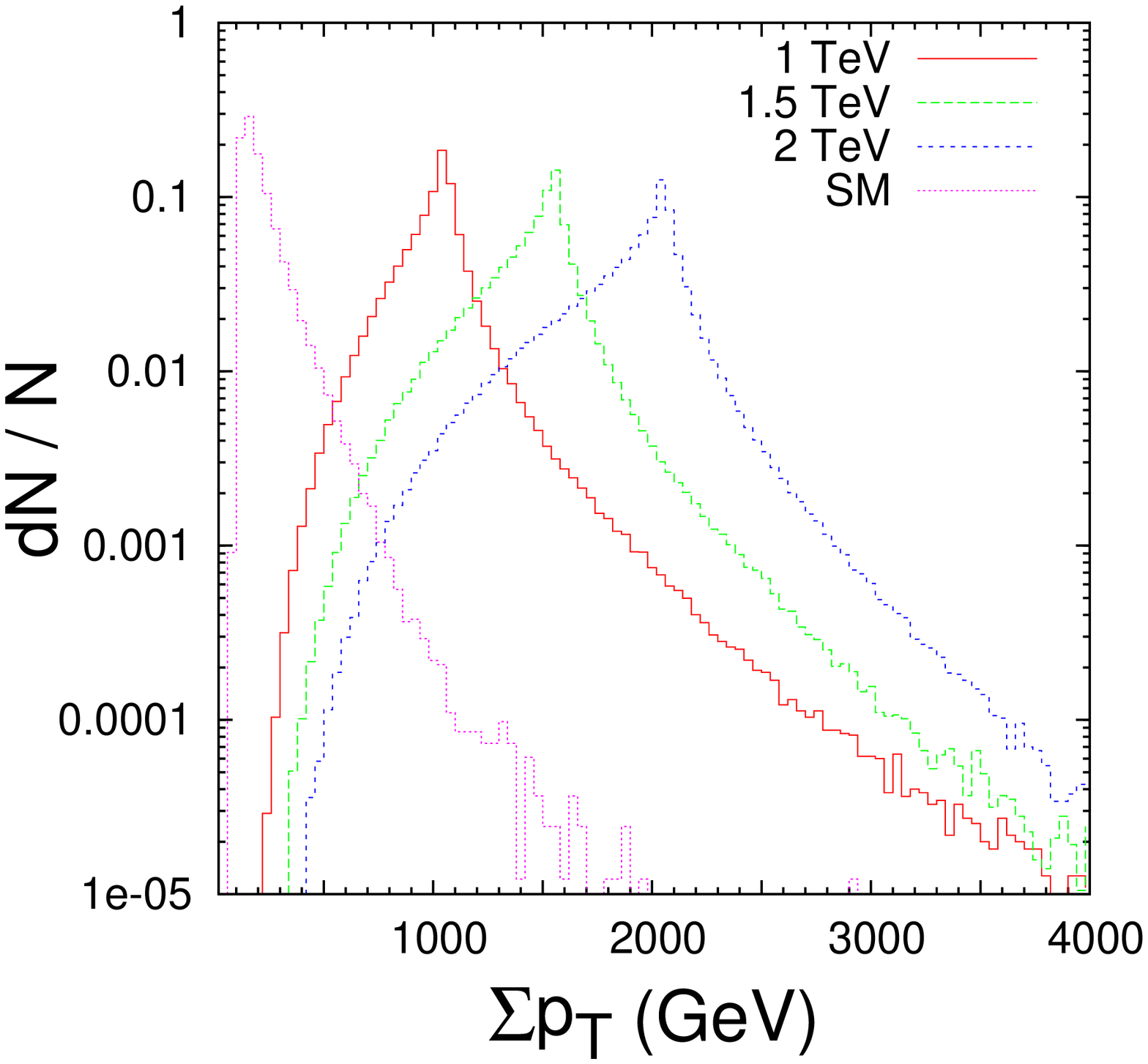}}
\caption{\small\sf Scalar $p_T$ distributions for the signal and the 
background. Choices of colors are the same as in Figure \ref{fig:ptg}.}
\label{fig:pts}
\end{figure}

The SM background is mostly composed of (a) the direct triphoton 
production i.e. where three photons are produced by SM processes, and 
(b) processes involving fake photons, in other words, where the detector 
incorrectly identifies particles which are not photons as photons.

In order to estimate the SM background we use {\tt MADGRAPH} with the 
same parton densities as for the signal case. At 14 TeV, the total 
cross-section for the direct triphoton production in the SM is $0.11$ 
pb. The most likely fake photon backgrounds are those due to the 
misidentification of jet or an electron as a photon We can thus 
subdivide the processes in case (b) as follows:

(i) \underline{Jet induced:} $jjj$, $jj\gamma$ $j\gamma\gamma$, and,

(ii) \underline{Electron induced:} $\gamma\gamma e^{\pm}$, $\gamma 
e^+e^-$ and $e^{\pm} e^+e^-$.

(iii)\underline{Both jet and electron induced:} $j e^+ e^-$, $jje^\pm$, 
$j\gamma e^\pm$.

We found that the bare cross-sections for these processes 
are $\sigma(jjj) =3.1\times 10^7$ pb, $\sigma(jj\gamma) = 4.4\times 
10^4$ pb and, $\sigma(j\gamma\gamma) = 99.7$ pb. With the fake 
probability for a jet to be identified as a photon, $f_{j\to\gamma} = 
1.1\times 10^{-3}$, as given in~\cite{Aad:2009wy}, the contributions due 
to these channels to the triphoton background will be $0.04$ pb, 
$0.05$ pb and $0.11$ pb respectively.

\begin{figure}
\centerline{
\includegraphics[angle=-90, width=0.6\textwidth]{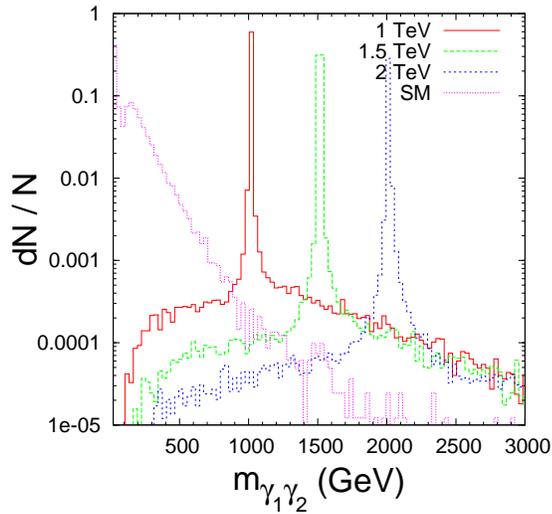}}
\caption{\small\sf Reconstructed graviton mass from the correct diphoton 
pair. Choices of colors are the same as in Figure \ref{fig:ptg}}
\label{fig:invm}
\end{figure}

Cross-sections for the processes involving electrons (and positrons) are 
$\sigma(e^{\pm}\gamma\gamma) = 0.05$ pb, $\sigma(e^+e^-\gamma) = 10.1$ 
pb, and, $\sigma(e^{\pm}e^+e^-) = 0.05$ pb. By assuming a fake rate 
similar to that of a jet the net contribution to the background from 
these processes will be $6.2\times 10^{-5}$ pb which is mostly due to 
the process involving one electron or a positron arising due to a $W$ 
production, where $W$ decays into an electron or a positron and the 
corresponding neutrino or anti-neutrino.

Contribution due to processes involving one or more jets and electrons 
(or positrons) is $2.5\times 10^{-5}$ pb which is also insignificant even 
though the bare rates are $290.1$ pb, $535.3$ pb and $19.9$ pb for $j e^+ 
e^-$, $jje^\pm$, $j\gamma e^\pm$ respectively.

The background form fake photons due to jets is the highest with a 
combined cross-section of $\simeq ~.2$ pb which is about two times 
larger than the direct SM triphoton background.

\subsection{Results and Graviton Reconstruction}

In Figures~\ref{fig:ptg},~\ref{fig:pts} and~\ref{fig:invm}, we present 
$p_T$, $\Sigma p_T$ and invariant mass $m_{12}$ distributions with basic 
cuts for three different values of graviton mass in case of signal, and, 
the net background due to direct tri-photons and faked backgrounds due 
to electorns (or positrons) and photons as discussed above. From 
Figures~\ref{fig:ptg}, it is clear that a demand of 
$p_{T_{\gamma_{1,2}}} > 100$ GeV and
$p_{T_{\gamma_3}} > 50$ GeV will ensure that the triphotons are indeed 
due to the a heavy graviton production. 
We employ further cuts on the 
scalar sum of photons transverse momenta, $\Sigma p_T$ and invariant 
mass of harder photons, $m_{12}$. Effects of these individual and 
combined cuts on the signal and background event rates for an LHC 
luminosity $\int {\cal L} dt = 300 $ fb$^{-1}$ is presented in 
Table~\ref{t:lhc}. As it clear from the Table~\ref{t:lhc} that, though a 
large fraction of the background is already eliminated at the basic 
level itself due to the demand that the missing transverse energy per 
event should not exceed $30$ GeV limit. Yet a cut on scalar sum of the 
transverse momenta of all the photons, $\Sigma p_T$ of about 550 GeV or 
alternatively, $m_{12}$ of about 800 GeV in addition to the basic and 
photon selection cuts reduces the background significantly. A 
combination of all these cuts reduces to background events to 14 while 
keeping signals events to 293, 213 and 105 respectively for the graviton 
mass 1, 1.5 and 2 TeV respectively for an integrated luminosity of 300 
fb$^{-1}$.

In order to reconstruct the graviton mass we use the following 
techniques: We first reconstruct all possible diphoton pairs with the 
$p_T$ ordered photons. Later we drop those pairs in each event which are 
nearer to each other, i.e., those with $|\Delta m_{ij}| < 20$ GeV 
where, $\Delta m_{ij} = m^a_{ij} - m^b_{ij}$ is the difference of 
reconstructed invariant masses. Clearly with this technique we are able 
to help a graviton mass with an uncertainty within $5-10$ per cent even 
in the presence of background as shown in Figure~\ref{fig:invm}.

\begin{figure}[htb]
\centerline{
\label{fig:ct1}\includegraphics[angle=-90, width=0.6\textwidth]{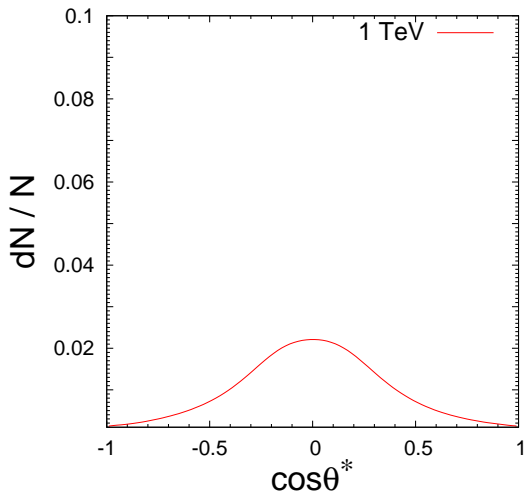}}
\caption{\small\sf Angular distribution of the decayed photon in the graviton 
rest frame. $m_G = 1$ TeV is assummed here.}
\label{fig:angle}
\end{figure}

\begin{table}[h]
\centering
\begin{tabular}{| c | c | c |c|c|c|}
\hline\hline
Cuts& SP1& SP2 & SP3 & SM Background\\\hline
Basic   &454&336&173&8516\\\hline
$p_{T_{\gamma_{1,2}}} > 100$ GeV,  $p_{T_{\gamma_3}} > 50$ GeV,  
 $\Sigma p_T > 550$ GeV&454&335&170&142\\
$p_{T_{\gamma_{1,2}}} > 100$ GeV,  $p_{T_{\gamma_3}} > 50$ GeV,  $m_{\gamma_1\gamma_2} > 800$ GeV&293&214&108&40\\\hline
 Combined&293&213&105&14\\
\hline\hline   
\end{tabular}
\caption{\small\sf Efficiency of cuts on signal and background triphoton 
events for an integrated LHC luminosity of $300$ fb$^{-1}$ }
\label{t:lhc}
\end{table}

Once we have reconstructed the graviton mass, the next task is to 
measure its spin in order to ensure that the photons are indeed due to 
decay of a RS graviton. In order to do so, we first identify the correct 
photon pair which yield the right graviton mass peak in a clean 
signature case. Next, we boost the photon pair into the rest frame of 
the graviton.  In this boosted frame we produce angular distribution of 
either of the photons as the two have nearly identical shape. We plot 
angular distribution for the first photon in Figure~\ref{fig:angle}.

\section{Summary and Conclusion}

We have studied triphoton signals in details. With a detailed study on 
background arising due to various SM processes including the processes 
where a photon can be faked by a jet or an electron or positron. We 
found that the processes is relatively clean and the background may be 
greatly reduced (see the Table~\ref{t:lhc}). We have also shown that the 
graviton mass reconstruction and its spin measurement is possible in 
this signals though we need to wait for 300 fb$^{-1}$ data in order to 
have sufficient statistics at the LHC.

\bigskip

{\large\bf Acknowledgments}

SKG thanks C Chen and J Cochran for valuable discussions on the SM 
background. This work was supported in part by a DOE grant under 
contract number DE-FG02-01ER41155.

\end{document}